\documentclass[aps,pra,twocolumn,groupedaddress,amsmath,amssymb,10pt]{revtex4}
\usepackage{epsfig}
\usepackage{amssymb}
\usepackage{amsmath}
\usepackage{graphicx}
\usepackage{amsfonts}
\usepackage{epsfig}
\usepackage{revsymb}
\usepackage{bm}
\usepackage{amsmath}
\usepackage{amssymb}
\usepackage{latexsym}
\usepackage{hyperref}
\usepackage{cleveref}

\begin{document}

\title{Superconductivity in a semiconductor doped with the resonance negative U-centers}
\author{Serguei N. Burmistrov and Leonid B. Dubovskii }
\affiliation{Moscow Institute of Physics and Technology, 141700 Dolgoprudny, Russia
\\
and 
\\
Kurchatov Institute, 123182 Moscow, Russia}
\begin{abstract} 
We study the effect of doping the semiconductor with the resonance negative U-centers upon its superconducting transition temperature. The attraction of electrons at the U-centers leads to the enhancement of transition temperature. On the contrary, the resonant scattering of conduction electrons along with their hybridization at the U-centers results in reducing the transition temperature. The largest effect on the superconducting transition temperature takes place if the resonance width is of the order of the energy level at the U-center. In the lack of the electron-phonon coupling under sufficiently low concentration of U-centers the superconducting transition is entirely absent in spite of the presence of attraction between electrons at U-centers.  
\end{abstract}
\maketitle

\section{Introduction}
\par
So far, noticeable attention has been paid for studying the superconductivity in the semiconductor systems with the local resonance interaction between electrons at the U-centers. Most of interest is the case when the interaction between electrons occupying the U-center corresponds to attraction between them. In this case we may expect that the pair fluctuations of electrons between the delocalized states in the conduction band and the resonant quasi-local states at the U-centers should favor an existence of superconductivity 
and increase the superconducting transition temperature. Such pair fluctuations result from the hybridization of conduction electrons with the resonant electron states at the U-center. Note here that the hybridization of electron states will produce some finite width $\Gamma$ of local states.  
\par
The local resonance states with the negative energy $U$ of interaction for the opposite spin electrons have been introduced by Anderson~\cite{Anderson}.  The origin of negative U-centers is considered as a presence of some type of internal degree of freedom, chemical bonds~\cite{Kastener}, appropriate region of electron-phonon coupling~\cite{Cohen}, bivalent fluctuating molecules~\cite{Schuttler}, and existence of Jahn-Teller centers~\cite{Bersuker}. According to the authors~\cite{Andreev,Lykov}, there is a reason  to assume an existence of negative U-centers associated with the In or Tl dopants in superconducting semiconductor PbTe. The doped semiconductor PbTe$_{1-x}$Tl$_x$ can thus represent the system where an existence of superconductivity is related with the presence of resonance impurity states at the U-centers. 
Superconductivity is observed in the semiconducting compounds GeTe \cite{Heim} and SnTe \cite{Heim1}.
\par 
The unusual behavior of superconducting transition temperature has been observed in a series of mixed barium-strontium Ba$_x$Sr$_{1-x}$TiO$_3$ and calcium-strontium Ca$_x$Sr$_{1-x}$TiO$_3$ titanate ceramic specimen introduced by partial substitution of Ca or Ba into SrTiO$_3$. The significant result \cite{Schooley,Schooley1} is that for the electron concentration smaller than 10$^{20}$\, cm$^{-3}$, the transition temperature has been enhanced. At higher carrier concentrations the transition temperature decreases rapidly. The superconducting transition temperature $T_c$ peaks to 0.45~K around a carrier density 10$^{20}$\, cm$^{-3}$ \cite{Schooley2,Koonce,Lin}. The fascinating problems of superconductivity in doped SrTiO$_3$ can be found in the recent reviews \cite{Gastiasoro, Collignon}. 
The effect of the electron-phonon coupling on superconductivity in Nb-doped SrTiO$_3$ is discussed in \cite{Park}.
 \par 
As will be shown below, local resonance states with the electron-electron attraction have a twofold effect on the superconducting transition temperature $T_c$. The attraction between electrons in the localized state promotes for the growth of superconducting transition temperature.  On the contrary, the resonant scattering of conduction electrons leads to decreasing the transition temperature as a result of their hybrydization with the localized states. The aim of the present work is an analysis for the behavior of the superconducting transition temperature in the doped semiconductor as a function of concentration of local resonance 
U-centers.

\section{Superconductor with the resonance U-centers}
\par 
We inspect the problem addressed by employing the BCS Hamiltonian augmented with the terms describing the resonant local states and transitions between conduction electrons and local electronic 
states:   
\begin{gather}
H=\sum\limits_{\bm{p}}\xi_{\bm{p}}a^{+}_{\bm{p}\alpha}a_{\bm{p}\alpha}-
g\sum\limits_{\bm{p},\bm{p}'}a^{+}_{\bm{p}'\uparrow}a^+_{\bm{-p}'\downarrow}a_{-\bm{p}\downarrow}a_{\bm{p}\uparrow} \nonumber
\\
+E_0\sum\limits_{m}b_{m\alpha}^+b_{m\alpha}+U\sum\limits_mn_{m\uparrow}n_{m\downarrow}
\label{equ1}
\\
+ \sum\limits_{\bm{p}, m}\bigl(V_{\bm{p},m}a_{\bm{p}\alpha}^+b_{m\alpha}
+V_{m,\bm{p}}b_{m\alpha}^+  a_{\bm{p}\alpha}   \bigr). \nonumber
\end{gather}
Here $a_{\bm{p}\alpha}^+$,  $a_{\bm{p}\alpha}$ are the creation and annihilation operators  of electrons with momentum $\bm{p}$ and spin projection $\alpha$, and $\xi_{\bm{p}}$  is the energy of conduction electron with momentum $\bm{p}$ taken from the Fermi energy $\mu$. The constant  $g>0$ expresses the effective attraction of conduction electrons with the opposite directions via phonons in the BCS model. 
\par 
The parameters specifying the U-centers and the coupling of the localized and delocalized electron states are as follows: $E_0$ is the level energy of an isolated local U-center, $n_{m\uparrow}=b_{m\uparrow}^+b_{m\uparrow}$ and $n_{m\downarrow}=b_{m\downarrow}^+b_{m\downarrow}$ are the density operator of localized electrons with the up and down spin at the site with number $m$, and $V$ is the amplitude of interaction between the conduction electron and the localized electron at the site. The summation over $m$ is performed over all the lattice sites occupied with the resonance centers.  
\par 
For comprehensibility, when analyzing the effect of localized states on the  superconducting transition temperature $T_c$, we below omit the effective BCS interaction for a while and restore it in the final answer. Let us write Hamiltonian $H$ \eqref{equ1} in the equivalent form with the aid of $\psi ^+(\bm{r})$ and $\psi (\bm{r})$-operators (hereafter $\hbar =1$)
\begin{gather}
H=\int \psi_{\alpha}^+(\bm{r})\biggl(-\frac{\nabla^2}{2m}-\mu\biggr)\psi _{\alpha}(\bm{r})d^3r\nonumber 
\\
+E_0\sum\limits_mb_{m\alpha}^+b_{m\alpha}+\frac{U}{4}\sum\limits_mb_{m\alpha}^+b_{m\beta}^+L_{\alpha\beta\gamma\delta}b_{m\delta}b_{m\gamma}+ \label{equ2}
\\
\sum\limits_m\!\int\! d^3r\biggl[V(\bm{r}\!-\!\bm{R}_m)\psi_{\alpha}^+(\bm{r})b_{m\alpha}\!+\!V(\bm{r}\!-\!\bm{R}_m)b_{m\alpha}^+\psi_{\alpha}(\bm{r})\biggr]. \nonumber 
\end{gather}
Here the spin operator  is antisymmetric $L_{\alpha\beta\gamma\delta}=\delta_{\alpha\gamma}\delta_{\beta\delta}-\delta_{\alpha\delta}\delta_{\beta\gamma}$. To simplify the description of the phenomenon, we take the interaction corresponding to hybridization as point-like, i.e.  $V(\bm{r}-\bm{R}_m)=V\delta (\bm{r}-\bm{R}_m)$, and thus $V_{\bm{p},m}=V\exp( -i\bm{pR}_m)$. In addition, we assume that the resonance centers are distributed randomly in the matrix of semiconductor. This allows us to employ the routine approach \cite{Abrikos} for the metals with impurities. In particular, such approach \cite{Ting} has been used for analyzing the effect due to tunneling the conduction electrons into negative U-centers at the disordered metal-semiconductor interfaces on the superconductivity of metal-semiconductor eutectic Al-Si, Al-Ge, and Be-Si alloys. Note here that this situation with impurities is also not artificial for the superconducting semiconductor PbTe$_{1-x}$Tl$_x$ \cite{Lykov}. 
\par 
The presence of a hybridization term in Hamiltonian $H$ \eqref{equ1}, which describes the mixing of conduction electrons with the local states, will lead us not only to the appearance of effective electron-electron attraction, but also to other obscure effects. First, the width of the localized states together with the damping in the dispersion law of conduction electrons will appear. Second,  the renormalization of interaction $U$  itself between localized states emerges as well. Before investigating the effective attraction between conduction electrons, let us first consider the last two effects. 

\section{Width of localized states and the damping of conduction electrons}
\par 
As usual, let us introduce temperature Matsubara Green functions \cite{Lifshits}  for the conduction electrons and localized states 
\begin{eqnarray}
G_{\alpha\beta}(x, x')=-\langle T_{\tau}\psi_\alpha (x)\psi_{\beta}^+(x'), \quad x=(\bm{r},\tau); \nonumber
\\
g_{\alpha\beta}(x, x')=-\langle T_{\tau}b_\alpha (x)b_{\beta}^+(x'), \quad x=(m,\tau). 
\label{equ3}
\end{eqnarray}
The initial Green functions corresponding to the free Hamiltonian $H^{(0)}$
\begin{eqnarray}
H^{(0)}\!=\!\!\!\int\! \psi_{\alpha}^+(\bm{r})\biggl(\!-\frac{\nabla^2}{2m}-\mu\!\biggr)\psi _{\alpha}(\bm{r})d^3r +\! E_0\!\sum\limits_mb_{m\alpha}^+b_{m\alpha} 
\label{equ4}
\end{eqnarray}
can straightforwardly be found. Accordingly, the initial Green functions for conduction electrons and localized states are equal to 
\begin{equation}
G^{(0)}=\frac{1}{i\omega-\xi_{\bm{p}}} \quad\text{and}\quad g^{(0)}=\frac{1}{i\omega -E_0} 
\label{equ5}
\end{equation}
where the odd Matsubara frequency is $\omega =\pi T(2n+1)$, $n=0, \pm 1, \pm 2, \ldots$ and  $\xi_{\bm{p}}=\bm{p}^2/2m -\mu$. 
\par 
In general, the Green functions for the metal with the resonance centers should depend on the two momenta due to spatial inhomogeneity of the medium. However, after averaging over spatial configurations of  random locations of resonance centers,  the Green functions become diagonal in the momenta. The graphical illustration for the equations is shown in the figure below.  
\begin{figure}[ht]
\begin{center}
\includegraphics[scale=0.7]{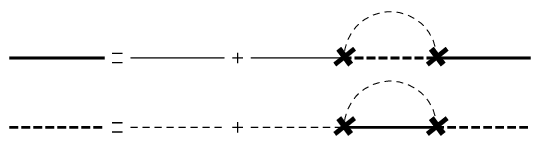}
\caption{The solid lines denote the Green functions $G(\bm{p},\omega)$ of conduction electrons. The dashed lines are the Green functions $g(\omega)$ of localized states.  The cross implies the scattering of conduction electrons at the resonance U-center.}
\label{gr1}
\end{center}
\end{figure}
In Fig.~\ref{gr1} the Green functions of conduction electrons are shown with the solid lines and the localized states are done with the dashed ones. The scattering of conduction electrons with the resonance U-center is indicated with the cross. Accordingly, the equations for the renormed Green functions are, as follows:  
\begin{eqnarray}
G(\bm{p}, \omega)=G^{(0)}(\bm{p}, \omega)+G^{(0)}(\bm{p}, \omega)\, cV^2g(\omega)G(\bm{p}, \omega), \nonumber
\\
g(\omega)=g^{(0)}(\omega)  +g^{(0)}(\omega)\,V^2\int\frac{d^3p}{(2\pi)^3}G(\bm{p}, \omega)g(\omega) 
\label{equ6}
\end{eqnarray}
where $c$ is the concentration of resonance U-centers. We get the both Green functions from these two equations above
\begin{gather}
g(\omega)=\frac{1}{i\omega -E+i\Gamma\text{sgn}\,\omega}, \nonumber
\\
G(\bm{p},\omega)=\biggl[i\omega\frac{cV^2\varkappa (\omega)}{E^2+\omega ^2\varkappa^2(\omega)}+\frac{cV^2E}{E^2+\omega^2\varkappa^2(\omega)}-\xi_{\bm{p}}\biggr]^{-1}, \nonumber 
\\ 
 \varkappa(\omega )=1+\frac{\Gamma}{\vert\omega\vert}. 
\label{equ7}
\end{gather}
Here $\Gamma=\pi N(0)V^2$ where $N(0)$ is the density of electron states at the Fermi surface with the fixed spin direction. The quantity $\Gamma$ is the width of the quasi-local state as a result of hybridization processes. 
\par 
The electron Green function has the well-known structure in the vicinity of the Fermi surface   
\begin{equation}
G(\bm{p},\omega)=\biggl[i\omega \biggl(1+\frac{1}{2\tau \vert\omega\vert}\biggr)-\xi_{\bm{p}}\biggr]^{-1} 
\label{equ8}
\end{equation} 
where we have involved an insignificant $\mu\rightarrow \mu +\delta\mu$ renormalization of chemical potential as $\delta\mu =-cV^2E(E^2+\Gamma^2)^{-1}$. The magnitude for the damping of electron states  is equal to  
\begin{equation}
\frac{1}{2\tau}=cV^2\frac{\Gamma}{E^2+\Gamma^2}
\label{eq9}
\end{equation}
and proportional to the concentration $c$ of resonance centers and to fourth power of hybridization potential  $V$ since  $\Gamma\sim V^2$. We have the density of local states at the Fermi surface from the above Eq.~\eqref{equ7} 
\begin{equation}
\nu _b(0)=\frac{1}{\pi}\frac{\Gamma}{E^2+\Gamma^2}.  
\label{equ10}
\end{equation}
The density of local states $\nu _b(0)$ vanishes together with the hybridization potential $V$.

\section{Renormalization of local interaction and energy of local state}
\par
\begin{figure}[ht]
\begin{center}
\includegraphics[scale=0.7]{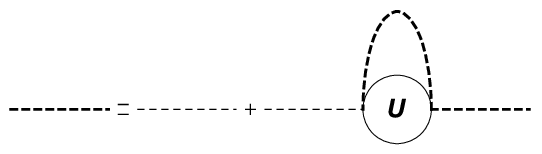}
\caption{The thin dashed line marks the probed Green function $g^{(0)}(\omega)$ and the solid dashed line does the renormalized Green function $g(\omega)$ of localized states. The empty circle denotes interaction  $U$ between electrons in the local state.}
\label{gr3}
\end{center}
\end{figure}

 
Let us start from considering the effect of interaction $U$ between electrons in the local state on the position of the energy level $E_0$ of local state. The trivial equation to find the Green function   $g(\omega)$ of localized states is illustrated in Fig.~\ref{gr3}.
The circle means the interaction $U$ between electrons in the local state.  Accordingly, the equation for the Green function $g(\omega)$ reads  
\begin{gather}
g(\omega)=g^{(0)}(\omega) +g^{(0)}(\omega)\Sigma\, g(\omega), \nonumber
\\
\Sigma =UT\sum\limits_{\omega_1}g(\omega_1). 
\label{equ11}
\end{gather}
Solving this equation, we have expression~\eqref{equ5} for the Green function $g(\omega)$ with the renormalized magnitude of energy in the local state
\begin{equation}
E=E_0+\frac{U}{2}n. 
\label{equ12}
\end{equation}
Here $n$ is the mean density of the localized electrons at the resonance center. 
\par
\begin{figure}[ht]
\begin{center}
\includegraphics[scale=0.95]{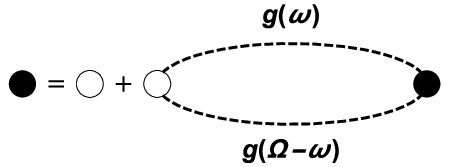}
\caption{The Green functions $g(\omega)$ and $g(\Omega-\omega)$ of localized states are specified by the dashed lines. The empty circle means the probed interaction $U$ of electrons at the U-center. The shaded circle indicates the renormalized interaction $\Gamma (\Omega)$ at the U-center.}
\label{gr2}
\end{center}
\end{figure}

 
An existence of hybrydization with the conduction electrons, resulted in width $\Gamma$ of localized level, will also entail the variation of interaction $U$ between electrons in the localized state at the same center. For the renormalized vertex, we have the following graphical equation in Fig.~\ref{gr2}. 
In Fig.~\ref{gr2} the renormalized interaction $\Gamma (\Omega)$ is shown by the shaded circle and the dashed lines denote the renormalized Green functions $g(\omega)$. The equations corresponding to the diagram are the following: 
\begin{gather}
\Gamma(\Omega)=U+U\Pi(\Omega)\Gamma (\Omega), \nonumber
\\
\Pi (\Omega)=-\frac{1}{2}T\sum\limits_{\omega} g(\omega)g(\Omega -\omega) 
\label{equ13}
\end{gather}
where $\Omega$ is a sum of the frequencies of the incoming particles. Solving these Eqs.~\eqref{equ13} and calculating $\Pi (\Omega)=\Pi _1(\Omega) +i\Pi _2(\Omega)$, we get 
\widetext
\begin{gather}
\Pi_1(\Omega)=\frac{1}{\pi}\frac{\Gamma}{(2E-\vert\Omega\vert)^2+4\Gamma^2}\biggl[\ln\frac{(\vert\Omega\vert-E)^2
+\Gamma^2}{E^2+\Gamma^2}
+\frac{2(\vert\Omega\vert-2E)^2+4\Gamma^2}{(\vert\Omega\vert-2E)\Gamma}\arctan\frac{E}{\Gamma}+\frac{4\Gamma}{\vert\Omega\vert -2E}\arctan\frac{E-\vert\Omega\vert}{\Gamma}\biggr], \nonumber
\\
\Pi_2(\Omega)=-\frac{\text{sgn}\,\Omega}{\pi}\frac{2\Gamma}{(\vert\Omega\vert -2E)^2+4\Gamma^2}
\biggl[\frac{\Gamma}{\Omega\vert -2E}\ln\frac{(\vert\Omega\vert -E)^2+\Gamma^2}{E^2+\Gamma^2}+\arctan\,\frac{E}{\Gamma}-\arctan\,\frac{E-\vert\Omega\vert}{\Gamma}\biggr].
\end{gather}
\widetext
\twocolumngrid
\par 
As calculating the superconducting transition temperature $T_c$, we will be interested in the case of zero total frequency $\Omega =0$.  In place of initial interaction magnitude $U$ the following parameter of effective interaction $U_{\text{eff}}$ emerges and equals  
\begin{equation}
U_{\text{eff}}=\frac{U}{1+\frac{U}{\pi E}\arctan\frac{E}{\Gamma}}. 
\label{equ14}
\end{equation}
  
\section{Resonance attraction between the conduction electrons}
\par
\begin{figure}[ht]
\begin{center} 
\includegraphics[scale=0.7]{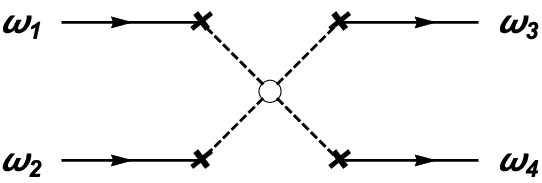}
\caption{The solid lines are related with the conduction electrons. The dashed lines correspond to the localized states. The crosses here denote the interaction of magnitude $V$. The circle is the resonance U-center. }
\label{gr5}
\end{center}
\end{figure}

Let us consider the following process.  A pair of conduction electrons collide with the same U-center, go  into the localized states, interact with each other, and after cross over again to the delocalized states of conduction electrons. Such process can be associated with the plot in Fig.~\ref{gr5}, 
representing the effective interaction between the conduction electrons. This effective interaction is completely analogous to the ordinary four-fermion one. As usual, averaging over the randomly spatial localization of U-centers, we obtain the relation for the effective interaction between the conduction electrons 
\begin{eqnarray}
\Gamma_{\alpha\beta\gamma\delta}(p_1,p_2,p_3,p_4)=(2\pi)^4\delta (p_1+p_2-p_3-p_4)  \nonumber
\\
\times L_{\alpha\beta\gamma\delta}\,\Gamma (\omega_1,\omega_2,\omega_3,\omega_4), \; p=(\bm{p},\omega) 
\label{15}
\end{eqnarray}
where $L_{\alpha\beta\gamma\delta}=\delta_{\alpha\gamma}\delta_{\beta\delta}-\delta_{\alpha\delta}\delta_{\beta\gamma}$ is the antisymmetric spin operator as in Eq.~\eqref{equ2}. As is expected, only the  electrons with the opposite spins interact and the vortex $\Gamma (\omega_1,\omega_2,\omega_3,\omega_4)$ is given by 
\begin{equation}
\Gamma (\omega_1,\omega_2,\omega_3,\omega_4)=cUV^4g(\omega_1)g(\omega_2)g(\omega_3)g(\omega_4). \label{equ16}
\end{equation}
\par 
Let us treat vertex $\Gamma(p_1,p_2,p_3,p_4)$ at frequencies $\omega _i=\xi (p_i)$ where ($i=1,2,3,4$). 
This vertex plays a key role for the amplitude of the quasiparticle-quasiparticle scattering at the Fermi surface. In other words, we need the vertex~\eqref{equ16} at zero frequencies, i.e.
\begin{equation}
\Gamma (0)=\Gamma (+0,-0,+0,-0)=\frac{cU}{\pi^2N^2(0)}\frac{\Gamma^2}{(E^2+\Gamma^2)^2}. \label{equ17}
\end{equation}
This amplitude $\Gamma (0)$ will be encountered in the course of calculating the superconducting transition temperature. 
\begin{figure}[ht]
\begin{center}
\includegraphics[scale=0.7]{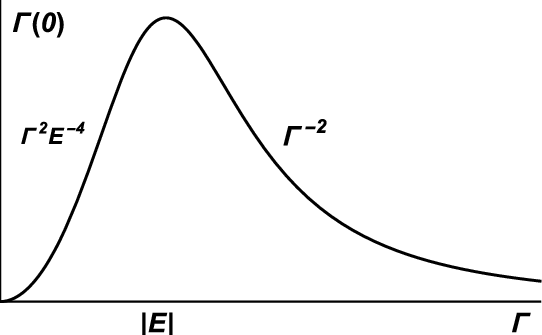}
\caption{Amplitude $\Gamma (0)$ as a function of the local level width $\Gamma$ with energy $E$.}
\label{Fig1}
\end{center}
\end{figure}
We can see from Eq.~\eqref{equ17} that the sign of effective interaction between the conduction electrons coincides with the sign of interaction $U$. 
\par 
In Fig.~\ref{Fig1} we show the amplitude $\Gamma (0)$ as a function of the width of local level with energy  $E$. Hence, the interaction is weak for the small width of the local level as well as for the very large width. The maximum magnitude of interaction reaches at $\Gamma\sim \vert E\vert$. Such behavior can readily be understood. For the large width $\Gamma$, the electrons in the localized state have no time to interact since their  lifetime at the U-center is short. In the opposite limit of small width $\Gamma$ the electrons are mostly located at the U-center and cannot run away.  

\section{Determination of the superconducting transition temperature}
\par 
In the previous section we have found the magnitude of effective interaction between the electrons. So, we can determine the superconducting transition temperature. Let us write equation for the vertex of conduction electrons in the Cooper channel.  
\begin{figure}[ht]
\begin{center}
\includegraphics[scale=0.5]{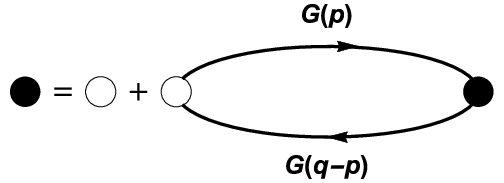}
\caption{The diagram for the vertex $\Gamma (q,p)$ in the Cooper channel. The solid lines are the Green functions. }
\label{gr4}
\end{center}
\end{figure}
The equation, which corresponds to the plot in Fig.~\ref{gr4} and determines the pole in the vertex,
has a usual form \cite{Lifshits}
\begin{equation}
1=-T\sum\limits_{\omega}\int\!\frac{d^3p}{(2\pi)^3}\Gamma(\Omega -\omega, \omega)G(\bm{p},\omega)G(-\bm{p},\Omega -\omega). 
\label{equ18}
\end{equation}
Here the temperature Matsubara Green function  
\begin{equation}
G^{-1}(\bm{p},\omega)=\Sigma '(\omega)+i\Sigma^{\prime\prime}(\omega)-\xi_{\bm{p}} 
\label{equ19}
\end{equation}
is determined with expression~\eqref{equ7}. The interaction $\Gamma (\Omega-\omega, \omega)$ is 
given by Eq.~\eqref{equ16}
\begin{equation}
\Gamma (\Omega-\omega, \omega)=cU_{\text{eff}}(\Omega)V^4 g^2(\Omega -\omega)g^2(\omega),  
\end{equation}
$c$ being the concentration of U-centers. 
\par 
We expect that the first pole in temperature $T$ occurs at zero total frequency $\Omega=0$. Accordingly, equation \eqref{equ18} crosses over into the equation determining the superconducting transition temperature $T_c$: 
\begin{eqnarray}
1=-T\sum\limits_{\omega}\int\frac{d^3p}{(2\pi)^3}\, cU_{\text{eff}}(0)V^4 \nonumber
\\
\times g^2(\omega)g^2(-\omega)
G(\bm{p},\omega)G(-\bm{p},-\omega).
\label{equ21}
\end{eqnarray}
Integrating over momentum $\bm{p}$ yields  
\begin{equation}
1=-cU_{\text{eff}}(0)V^4N(0)\pi T\sum\limits_{\omega}\frac{g^2(\omega)g^2(-\omega)}{\Sigma^{\prime\prime}(\vert\omega\vert )}. 
\label{equ22}
\end{equation}
Substituting the expressions from~\eqref{equ19} and \eqref{equ7}, we get the final equation for determining the transition temperature $T_c$ as a sum over the Matsubara frequencies $\omega =\pi T(2n+1)$ ($n=0, \pm 1,\pm 2\ldots$):
\begin{gather}
\frac{1}{\lambda_{\text{eff}}}=\pi T\sum\limits_{\omega}\biggl(\frac{E^2+\Gamma^2}{E^2+(\vert\omega\vert +\Gamma)^2}\biggr)^2 \times \nonumber
\\ 
\biggl(\vert\omega\vert\biggl[1+a\frac{E^2+\Gamma^2}{E^2+(\vert\omega\vert +\Gamma)^2}\biggl(1+\frac{1+a}{a}\frac{1}{\vert\omega\vert\tilde{\tau}}\biggr)\biggr]\biggr)^{-1}. 
\label{equ23}
\end{gather}
Here we have denoted  
\begin{gather}
\lambda_{\text{eff}}=\Gamma(0)N(0)=\frac{c\vert U_{\text{eff}}(0)\vert}{\pi^2N(0)}\frac{\Gamma^2}{(E^2+\Gamma^2)^2}, \nonumber
\\
a=c \frac{\nu_b(0)}{N(0)}=\frac{c}{\pi N(0)}\frac{\Gamma}{E^2+\Gamma^2}, \label{equ24}
\\
\frac{1}{\tilde{\tau}}=\Gamma\frac{a}{1+a}=\frac{1}{2\tau}\frac{1}{1+a}. \nonumber
\end{gather}
As is seen from \eqref{equ23},  the sum over frequency $\omega$ converges at the frequency of the order of $\Omega\sim \text{max}(E,\Gamma)$. Usually the magnitude $\Gamma$ is about 0.1 -- 1 eV and thus we analyze the relations~\eqref{equ24} in the limit $\Gamma\gg 2\pi T_c$. Then the equation~\eqref{equ23} simplifies and reduces to the well-known equation in the theory of superconductors with paramagnetic impurities \cite{Abrikos1,Gennes}:
\begin{equation}
\ln\frac{T_{c0}}{T_c}=\psi \biggl(\frac{1}{2}+\frac{1}{2\pi\tilde{\tau}T_c}\biggr)-\psi\biggl(\frac{1}{2}\biggr).  
\label{equ25}
\end{equation}
Here $\psi (x)$  is the logarithmic derivative of gamma function or digamma function. So, we have 
\begin{equation}
T_{c0}\approx\Omega \exp\biggl(-\frac{1}{\tilde{\lambda}_{\text{eff}}}\biggr), \quad \tilde{\lambda}_{\text{eff}}=\frac{\lambda_{\text{eff}}}{1+a}. 
\label{equ26}
\end{equation}
\par 
Unlike superconductors doped with paramagnetic impurities,  in our system the role of magnetic scattering time plays the time $\tilde{\tau}$ \eqref{equ24} characterizing the lifetime of conduction electrons at the Fermi surface \eqref{eq9}. Similar to superconductors with paramagnetic impurities such damping results in decreasing the superconducting transition temperature. For an existence of superconductivity under the lack of electron-phonon coupling, it is necessary to satisfy condition $\tilde{\tau}T_{c0}>1$.  The behavior of effective coupling constant in the exponent for the critical temperature $T_{c0}$ has been analyzed in the previous section. 
\par 
Provided that there is an additional electron-phonon coupling in the system, equations \eqref{equ18} and \eqref{equ21} are readily generalized by augmenting the electron-phonon coupling to the resonance interaction. As a result, the critical temperature $T_c$ is governed with Eq.~\eqref{equ25} but we should replace $\lambda _{\text{eff}}\rightarrow\lambda_{\text{eff}}+\lambda_{\text{e-ph}}$ in formula~\eqref{equ26}. We suppose here that the Debye frequency $\omega_D$ is comparable with frequency  $\Omega$. 
\par 
It is interesting to note that the sign of derivative $dT_c/dc\vert _{c=0}$ can be either positive or negative as a function of combination between electron-phonon constant $\lambda_{\text{e-ph}}$ and magnitudes of $U$ and $\Gamma$. In other words, the doping of negative U-centers into semiconductor may result in increasing the superconducting transition temperature $T_c$ as well as in its decrease. The similar conclusion has been observed in works \cite{Schuttler,Schooley2,Koonce,Lin}. 
\begin{figure}[ht]
\begin{center}
\includegraphics[scale=0.65]{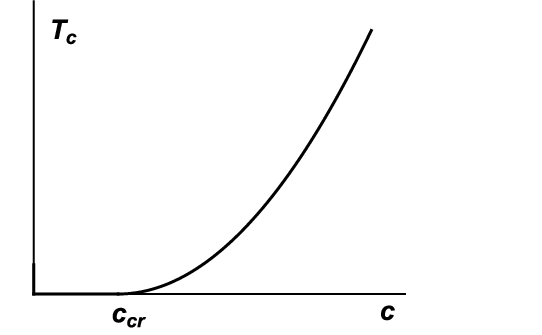}
\caption{The behavior of critical temperature $T_c$ vs. concentration $c$ of resonance U-centers. }
\label{Fig2a}
\end{center}
\end{figure}

\section{Summary} 
\par
\begin{figure}[ht]
\begin{center}
\includegraphics[scale=0.65]{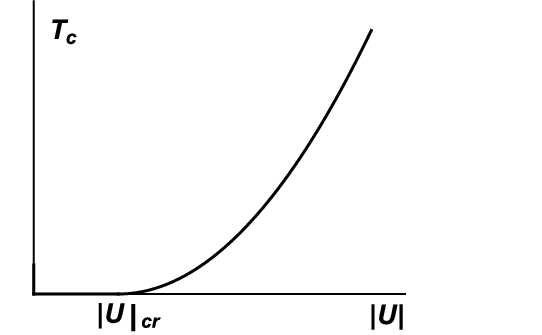}
\caption{The behavior of critical temperature $T_c$ vs. magnitude of interaction $U$ between electrons in the localized states.}
\label{Fig2b}
\end{center}
\end{figure}

In Figs. \ref{Fig2a} and \ref{Fig2b} we show the behavior of superconducting transition temperature  $T_c$ as a function of the U-center concentration $c$ and interaction magnitude $U$ in the system in the lack of electron-phonon coupling. As is seen from the figures, the critical temperature $T_c$ is nonzero at sufficiently large concentration of resonance U-centers and sufficiently large magnitude of attraction $U$ of electrons at the localized states at the U-centers. The reason for such behavior is that hybridization results simultaneously in appearing attraction between conduction electrons as well as the damping of these electrons due to their transition to another state. Thus the coherence of electron states in the Cooper pair breaks down. 
\par 
The growth of the superconducting transition temperature $T_c$ should be achieved by reducing the damping, i.e. frequency  
$\tilde{\tau}^{-1}$ determined in Eq. \eqref{equ24}.  
On the other side, since the attraction between electrons is also proportional to the U-center concentration and hybridization potential $V$, the decrease of frequency $\tilde{\tau}^{-1}$ entails decreasing the effective attraction between the conduction electrons. The latter does not allow us to increase the critical temperature $T_c$ of the superconducting transition. Undoubtedly, the growth of the transition temperature results from the increase of interaction magnitude at the resonance U-center.    
\par 
Superconductivity conserves even at the low U-center concentrations under occurrence of electron-phonon coupling. Doping the semiconductor with the resonance U-centers can bring us the growth of critical temperature as well as its lowering. For the small magnitudes of electron-phonon coupling constant 
$\lambda_{\text{e-ph}}$, the derivative $dT_c/dc_{\vert c=0}<0$  is negative and it requires $\lambda_{e-ph}>(\lambda_{e-ph})_{\text{min}}$ for the derivative $dT_c/dc_{\vert c=0}$ to be positive. From the physical point of view this fact means that the conduction electrons which occupy the resonance U-centers will drop the electron-phonon attraction while the electrons are at the U-center.  
\par 
Most optimum relation for increasing the superconducting transition temperature $T_c$ is provided that the level width $\Gamma$ is approximately equal to the level energy $E$. Under weak hybridization the conduction electrons occupy seldom the resonance U-center, whereas under strong hybridization the conduction electrons stay for a very short time to deliver the effective U-attraction. The presence of resonance negative U-centers in a semiconductor can significantly increase the critical temperature provided that their parameters are specially selected.   

\end{document}